\documentclass[aps,prd,superscriptaddress,showpacs,preprint,amsmath,amssymb]{revtex4}
\usepackage{graphicx, bm}
\usepackage[usenames]{color}

\begin{document}

\draft
\title{Probing model-independent limits on $W^+W^-\gamma$ triple gauge boson vertex at the LHeC and the FCC-he}

\author{ A. Guti\'errez-Rodr\'{\i}guez\footnote{alexgu@fisica.uaz.edu.mx}}
\affiliation{\small Facultad de F\'{\i}sica, Universidad Aut\'onoma de Zacatecas\\
         Apartado Postal C-580, 98060 Zacatecas, M\'exico.\\}

\author{M. K\"{o}ksal\footnote{mkoksal@cumhuriyet.edu.tr}}
\affiliation{\small Deparment of Optical Engineering, Sivas Cumhuriyet University, 58140, Sivas, Turkey.\\}

\author{A. A. Billur\footnote{abillur@cumhuriyet.edu.tr}}
\affiliation{\small Deparment of Physics, Sivas Cumhuriyet University, 58140, Sivas, Turkey.\\}

\author{ M. A. Hern\'andez-Ru\'{\i}z\footnote{mahernan@uaz.edu.mx}}
\affiliation{\small Unidad Acad\'emica de Ciencias Qu\'{\i}micas, Universidad Aut\'onoma de Zacatecas\\
         Apartado Postal C-585, 98060 Zacatecas, M\'exico.\\}

\date{\today}

\begin{abstract}

We study the cross-section of production of a single $W^-$ boson in association with a neutrino through the process
$e^-p \to e^-\gamma^*p \to \nu_eW^- p$. Additionally, we obtain the anomalous couplings $\Delta\kappa_\gamma$ and
$\lambda_\gamma$ of the $W^+W^-\gamma$ vertex at the Large Hadron Electron Collider (LHeC) and the Future Circular
Collider Hadron-Electron (FCC-he). The impact of the polarized beam due to the electron is also analyzed. Our best
limits for $\Delta\kappa_\gamma$ and $\lambda_\gamma$ at the $95\%$ C.L. are: $\Delta\kappa_\gamma = \pm 0.0017$,
$\lambda_\gamma = \pm 0.0053$ (unpolarized electron beam) and $\Delta\kappa_\gamma = \pm 0.0013$,
$\lambda_\gamma = \pm 0.0046$ (polarized electron beam) identifying the $W^-$ boson through the hadronic decay channel.
In addition, the $e^-\gamma^* \to \nu_eW^-$ collision is one of the clean, pure and simple process to probe the $W^+W^-\gamma$
coupling without the complications of QCD backgrounds.

\end{abstract}


\vspace{5mm}

\maketitle

\section{Introduction}

The Standard Model (SM) of elementary particle physics based on the gauge group $SU(2)_L\times U(1)_Y$, describes the electroweak
interactions as being mediated by the $\gamma$-photon, the $Z$-boson and the $W^{\pm}$-bosons \cite{SM1,SM2,SM3}.

The $W^\pm$-bosons are among the heaviest particles known of the SM. Although the properties of the $W^\pm$-bosons have been studied
for many years, measuring its mass, as well as its anomalous couplings with high precision remains a great challenge and an important
objective to prove the unification of the electromagnetic and weak interactions in the SM. High precision measurement of the
properties of these bosons has made these particles one of the most attractive particles for new physics research.

It is worth mentioning that it is very important to measure the masses of the $W^\pm$-bosons, as well as its anomalous couplings,
as accurately as possible to better understand the Higgs boson, refine the SM and test its global consistency.

The $\nu_eW^-$ production at the $e^-p$ colliders contains a lot of information on the existence of trilinear self-couplings among
$W^+W^-\gamma$ gauge bosons. These couplings in a consequence of the non-Abelian gauge structure of the SM, predict the existence
of the triple couplings $W^+W^-\gamma$ \cite{SM1,SM2,SM3}. The $W^+W^-\gamma$ triple gauge boson vertex is accessible at the present
and future colliders such as the Large Hadron Collider (LHC), the Large Hadron Electron Collider (LHeC), the Future Circular Collider
Hadron-Electron (FCC-he) and the Compact Linear Collider (CLIC) at CERN for the post LHC era.

Under these arguments, in this paper, we study and present our results on the cross-section of the process $e^-p \to e^-\gamma^*p
\to \nu_eW^- p$. In addition, we obtain model-independent limits on the anomalous electromagnetic couplings $\Delta\kappa_\gamma$ and
$\lambda_\gamma$ of the $W^+W^-\gamma$ vertex for the high-energies of the center-of mass energies $\sqrt{s}=1.30, 1.98, 7.07, 10\hspace{0.8mm}
{\rm TeV}$ and high-luminosities ${\cal L}=10-1000\hspace{0.8mm}{\rm fb^{-1}}$ of the LHeC and the FCC-he \cite{FCChe,Fernandez,Fernandez1,Fernandez2,Huan,Acar}. We consider unpolarized and polarized electron beam.

A summary of experimental and phenomenological limits at $95\%$ C.L. on the anomalous triple gauge boson couplings
$\Delta\kappa_\gamma$ and $\lambda_\gamma$ from the present and future colliders are given in Table I of Ref. \cite{Billur}.
See Refs. \cite{Baur0,Hagiwara,Hagiwara1,Nachtmann,Sahin0,Cakir,Seyed,Ari,Atag,Atag1,Sahin,Papavassiliou,Choudhury,Chapon,Ellis}
for other limits on the anomalous $W^+W^-\gamma$ coupling in different contexts.

This work is organized as follows: In Sect. II we give an overview of the operators in our effective Lagrangian. In Sect. III we derive
limits on the anomalous couplings $\Delta\kappa_\gamma$ and $\lambda_\gamma$ at the LHeC and the FCC-he. In Sect. IV we present our conclusions.

\section{The triple gauge boson vertex $W^+W^-\gamma$ with anomalous contribution}

An appropriate model-independent context for describing possible new physics effects is based on effective Lagrangian. In this context,
all the heavy degrees of freedom are integrated out to obtain effective interactions between the SM particles. This is justified
since the related observables have so far not shown any significant deviation from the SM predictions.

We start from the effective Lagrangian formalism to study the process $e^-p \to e^-\gamma^*p \to \nu_eW^- p$, as well as to determine
limits on the anomalous couplings $\Delta\kappa_\gamma$ and $\lambda_\gamma$. In this regard, our starting point is the effective
Lagrangian ${\cal L}_{eff}$ for the $W^+W^-\gamma$ interaction of the photon and the gauge bosons with operators up to mass dimension-six.
Then ${\cal L}_{eff}$ can be expanded as:

\begin{equation}
{\cal L}_{eff}={\cal L}^{(4)}_{SM} + \sum_i \frac{C^{(6)}_i}{\Lambda^2}{\cal O}^{(6)}_i + {\rm h.c.},
\end{equation}

\noindent where ${\cal L}^{(4)}_{SM}$ denotes the renormalizable SM Lagrangian and the non-SM part contains ${\cal O}^{(6)}_i$ the
gauge-invariant operators of mass dimension-six. The index $i$ runs over all operators of the given mass dimension. The mass scale
is set by $\Lambda$, and the coefficients $C_i$ are dimensionless parameters, which are determined once the full theory is known.

Thus effective Lagrangian relevant to our analysis of $\Delta\kappa_\gamma$ and $\lambda_\gamma$ is given by:

\begin{equation}
{\cal L}_{eff}=\frac{1}{\Lambda^2}\Bigl[C_W{\cal O}_W + C_B{\cal O}_B + C_{WWW}{\cal O}_{WWW} + \mbox{h.c.}\Bigr],
\end{equation}

\noindent with

\begin{eqnarray}
{\cal O}_W&=&\bigl( D_\mu\Phi \bigr)^{\dagger}\hat W^{\mu\nu}\bigl( D_\nu\Phi \bigr),\\
{\cal O}_B&=&\bigl( D_\mu\Phi \bigr)^{\dagger}\hat B^{\mu\nu}\bigl( D_\nu\Phi \bigr),\\
{\cal O}_{WWW}&=&Tr\bigl[\hat W^{\mu\nu}\hat W^\rho_\nu\hat W_{\mu\rho}\bigr],
\end{eqnarray}

\noindent where $D_\mu$ is the covariant derivative, $\Phi$ is the Higgs doublet field and $\hat B_{\mu\nu}$, and $\hat W_{\mu\nu}$
are the $U(1)_Y$ and $SU(2)_L$ gauge field strength tensors. The coefficients of these operators $C_W/\Lambda^2$, $C_B/\Lambda^2$,
and $C_{WWW}/\Lambda^2$, are zero in the SM.

With this methodology, the effective Lagrangian for describing the $W^+W^-\gamma$ coupling can be parameterized as \cite{,Hagiwara,Gaemers}:

\begin{equation}
{\cal L}_{WW\gamma}=-ig_{WW\gamma} \Bigl[g^\gamma_1(W^\dagger_{\mu\nu} W^\mu A^\nu - W^{\mu\nu} W^\dagger_\mu A_\nu )
+\kappa_\gamma W^\dagger_{\mu} W_\nu A^{\mu\nu} + \frac{\lambda_\gamma}{M^2_W} W^\dagger_{\rho\mu} W^\mu_\nu A^{\nu\rho} \Bigr],
\end{equation}

\noindent where $g_{WW\gamma}=e$, $V_{\mu\nu}=\partial_\mu V_\nu -\partial_\nu V_\mu$ with $V_\mu= W_\mu, A_\mu$. The couplings
$g^\gamma_1$, $\kappa_\gamma$ and $\lambda_\gamma$ CP-preserving, and in the SM, $g^\gamma_1=\kappa_\gamma=1$ and $\lambda_\lambda=0$
at the tree level.

From Eq. (2), the operators of dimension-six are related to the anomalous triple gauge boson couplings as \cite{Rujula,Hagiwara1,Hagiwara2}:

\begin{equation}
\kappa_\gamma= 1+ \Delta\kappa_\gamma,
\end{equation}

\noindent with

\begin{eqnarray}
\Delta\kappa_\gamma&=&C_W + C_B,\\
\lambda_\gamma&=&C_{WW}.
\end{eqnarray}

From the effective Lagrangian given in Eq. (6), the Feynman rule for the anomalous $W^+W^-\gamma$ vertex function the most general
CP-conserving and that is consistent with gauge and Lorentz invariance of the SM is given by \cite{Hagiwara}:

\begin{eqnarray}
\Gamma^{WW\gamma}_{\mu\nu\rho}&=&e\Bigl[g_{\mu\nu}(p_1-p_2)_\rho + g_{\nu\rho}(p_2-p_3)_\mu + g_{\rho\mu}(p_3-p_1)_\nu
+\Delta\kappa_\gamma \Bigl( g_{\rho\mu}p_{3\nu}- g_{\nu\rho}p_{3\mu}\Bigr )  \nonumber\\
&+&\frac{\lambda_\gamma}{M^2_W}\Bigl(p_{1\rho}p_{2\mu}p_{3\nu} - p_{1\nu}p_{2\rho}p_{3\mu}
- g_{\mu\nu}(p_2\cdot p_3 p_{1\rho}-p_3\cdot p_1 p_{2\rho}) \nonumber\\
&-& g_{\nu\rho}(p_3\cdot p_1 p_{2\mu}-p_1\cdot p_2 p_{3\mu})
- g_{\mu\rho}(p_1\cdot p_2 p_{3\nu}-p_2\cdot p_3 p_{1\nu}) \Bigr)\Bigr],
\end{eqnarray}

\noindent where the first three terms in Eq. (10) corresponds to the SM couplings, while the terms with $\Delta\kappa_\gamma$
and $\lambda_\gamma$ give rise to the anomalous triple gauge boson couplings.

Different searches on these anomalous $W^+W^-\gamma$ couplings $\Delta\kappa_\gamma$ and $\lambda_\gamma$ were performed by the LEP,
the Tevatron and the LHC experiments, as shown in Table I of Ref. \cite{Billur}.

\section{Cross-section of the process $e^-p \to e^-\gamma^*p \to \nu_eW^- p$ and limits on the anomalous couplings
$\Delta\kappa_\gamma$ and $\lambda_\gamma$}

\subsection{Cross-section of the process $e^-p \to e^-\gamma^*p \to \nu_eW^- p$ at the LHeC and the FCC-he}

The LHeC and the FCC-he are proposed, designed and planned colliders to carry out $e^{-}p$ collisions at center-of-mass energies
$\sqrt{s}=1.30, 1.97, 7.07$ and 10 TeV, that is to say with a four main stage research region
\cite{FCChe,Fernandez,Fernandez1,Fernandez2,Huan,Acar}. The $e^{-}p$ colliders can also be operated as $e^-\gamma^*$,
$\gamma^* p$ and $\gamma^*\gamma$ collider. This enables the investigation of the $e\gamma^{*}$ interactions where the emitted
quasi-real photon $\gamma^{*}$ is scattered with small angles from the beam pipe of $e^{-}$ \cite{Ginzburg,Ginzburg1,Brodsky,Budnev,Terazawa,Yang}.
Since these photons have a low virtuality, they are almost on the mass shell. These processes can be described by the Equivalent
Photon Approximation (EPA) \cite{Budnev,Baur1,Piotrzkowski}, using the Weizsacker-Williams Approximation (WWA). The EPA has a lot
of advantages such as providing the skill to reach crude numerical predictions via simple formulae. Furthermore, it may principally
ease the experimental analysis because it enables one to directly achieve a rough cross-section for $e^-\gamma^{*} \to X$ process
via the examination of the main process $e^{-}p\rightarrow e^{-} X p$ where X represents objects produced in the final state. The
production of high mass objects is particularly interesting at the $e^-p$ colliders and the production rate of massive objects is
limited by the photon luminosity at high invariant mass while the $e\gamma^{*}$ process at the $e^-p$ colliders arises from quasi-real
photon emitted from the incoming beams. In many studies, new physics investigations are examined by using the EPA \cite{Abulencia,Aaltonen1,Aaltonen2,Chatrchyan1,Chatrchyan2,Abazov,Chatrchyan3,Inan,Inan1,Inan2,Sahin1,Atag2,Sahin2,Sahin4,Senol,
Senol1,Fichet,Sun,Sun1,Sun2,Senol2}.

Another very important element in our study corresponds to the impact of the polarization of the electron beam. About this,
in the baseline LHeC and FCC-he design, the electron beam can be polarized up to $\pm ±80\%$. By selecting different beam polarizations
it is possible to enhance or suppress different physical processes. In the particular case of the process
$e^-p \to e^-\gamma^*p \to \nu_eW^- p$, the chiral nature of the weak coupling to fermions can result in significant possible
enhancements in $\nu_eW^-$ production. Starting from this, the polarized $e^-$ beam combined with the clean experimental
environment provided by the LHeC and the FCC-he will allow to improve strongly the potential of searches for the $W^+W^-\gamma$ triple gauge
boson vertex. With these arguments, we consider polarized electron beam in our study. The expression for the total cross-section
for an arbitrary degree of longitudinal $e^-$ beam polarization is given by \cite{XiaoJuan}:

\begin{eqnarray}
\sigma_{e^-_r}= \sigma_{e^-_0}\cdot (1-P_{e^-_r}),  \hspace{7mm}
\sigma_{e^-_l} + \sigma_{e^-_r} &=& 2\sigma_{e^-_0},
\end{eqnarray}

\noindent where $\sigma_{e^-_r}$, $\sigma_{e^-_l}$ and $\sigma_{e^-_0}$ represent the right, left and without electron
beam polarization, respectively and $P_{e^-}$ is the polarization degree of the electron.

The schematic diagram corresponding to the process $e^-p \to e^-\gamma^*p \to \nu_eW^- p$ is given in Fig. 1. While the representative
leading order Feynman diagrams for the subprocess $\gamma^*e^- \to \nu_e W^-$ are depicted in Fig. 2. We based our calculations on
electron-photon fluxes through the subprocess $e^-\gamma^* \to  \nu_e W^- $. In addition, it is evident the contribution of elastic process
with an intact proton in the final state.

Finally, the total cross-section is obtained by folding the elementary cross-section with the photon distribution function:

\begin{eqnarray}
\sigma(e^-p \to e^-\gamma^* p \to \nu_e W^- p)=\int f_{\gamma^*}(x)\hat{\sigma}( e^-\gamma^* \to \nu_e W^-) dx,
\end{eqnarray}

\noindent where $\hat \sigma( e^-\gamma^* \to \nu_e W^-)$ is the cross-section for the reaction $e^-\gamma^* \to \nu_e W^-$,
and the distribution function $f_{\gamma^*}(x)$ of the EPA photons which are emitted by proton is given by \cite{Budnev,Belyaev}:

\begin{eqnarray}
f_{\gamma^*}(x)=\frac{\alpha}{\pi E_{p}}\{[1-x][\varphi(\frac{Q_{max}^{2}}{Q_{0}^{2}})-\varphi(\frac{Q_{min}^{2}}{Q_{0}^{2}})],
\end{eqnarray}

\noindent where $x=E_{\gamma}/E_{p}$ and $Q^2_{max}$ is the maximum virtuality of the photon. For our calculations, we use
$Q^2_{max}=2\hspace{0.8mm}GeV^2$. The minimum value of the $Q^2_{min}$ is:

\begin{eqnarray}
Q_{min}^{2}=\frac{m_{p}^{2}x^{2}}{1-x}.
\end{eqnarray}

From Eq. (13), the function $\varphi$ is given by:

\begin{eqnarray}
\varphi(\theta)=&&(1+ay)\left[-\textit{In}(1+\frac{1}{\theta})+\sum_{k=1}^{3}\frac{1}{k(1+\theta)^{k}}\right]
+\frac{y(1-b)}{4\theta(1+\theta)^{3}} \nonumber \\
&& +c(1+\frac{y}{4})\left[\textit{In}\left(\frac{1-b+\theta}{1+\theta}\right)+\sum_{k=1}^{3}\frac{b^{k}}{k(1+\theta)^{k}}\right],
\end{eqnarray}

\noindent where explicitly $y$, $a$, $b$ and $c$ are as follows:

\begin{eqnarray}
y=\frac{x^{2}}{(1-x)},
\end{eqnarray}

\begin{eqnarray}
a=\frac{1+\mu_{p}^{2}}{4}+\frac{4m_{p}^{2}}{Q_{0}^{2}}\approx 7.16,
\end{eqnarray}

\begin{eqnarray}
b=1-\frac{4m_{p}^{2}}{Q_{0}^{2}}\approx -3.96,
\end{eqnarray}

\begin{eqnarray}
c=\frac{\mu_{p}^{2}-1}{b^{4}}\approx 0.028.
\end{eqnarray}

In order to perform these calculations in an efficient way, we used a numerical method. The numerical integration
is performed using the CalcHEP packages \cite{Belyaev}.

The present LHeC and the FCC-he are planned to generate $e^-p$ collisions at energies from 1.30 TeV to 10 TeV \cite{Inan,Sahin}.
The LHeC is a suggested deep inelastic electron-nucleon scattering machine which has been planned to collide electrons with an
energy from 60 GeV to possibly 140 GeV, with protons with an energy of 7 TeV.  In addition, FCC-he is designed electrons with
an energy from 250 GeV to 500 GeV, with protons with an energy of 50 TeV. The LHeC and the FCC-he physics programs will enable
fundamentally new insights beyond the capabilities of the LHC for the anomalous coupling $W^+W^-\gamma$. In addition, the flexibility
and large accessible energy range provides a wide range of possibilities to measure the new physics using very different approaches.

The high-luminosity and the low backgrounds of QCD give access to the process $e^-p \to e^-\gamma^*p \to \nu_eW^- p$ at all
energies. Furthermore, the clean experimental environment and the good knowledge of the initial state allow precise measurements
of the cross-section of the $\nu_eW^- $ signal, as well as of $\Delta\kappa_\gamma$ and $\lambda_\gamma$, respectively.

Fig. 3 shows the total cross-sections of the process $e^-p \to e^-\gamma^* p \to \nu_e W^- p$ as a function of $\Delta \kappa_\gamma$
for center-of-mass energies of $\sqrt{s}=1.30, 1.98, 7.07, 10\hspace{0.8mm}{\rm TeV}$ at the LHeC and the FCC-he. The mechanism
$e^-p \to e^-\gamma^* p \to \nu_e W^- p$ is dominant at $\sqrt{s}=10\hspace{0.8mm}{\rm TeV}$ reaching a cross-section of $20\hspace{0.8mm}
{\rm pb}$. A similar study on the cross-sections of the process $e^-p \to e^-\gamma^* p \to \nu_e W^- p$ as a function of $\lambda_\gamma$
is presented in Fig. 4. In this case, the cross-section obtained is of $5\hspace{0.8mm}{\rm pb}$ at center-of-mass energy of
$\sqrt{s}=10\hspace{0.8mm}{\rm TeV}$. In general, this process can be identified at all the energy stages of the LHeC and the FCC-he.

From Fig. 3 (and similarly in Fig. 7), in the case of 1.30 and 1.98 TeV where the center-of-mass energies are relatively low,
there is an asymmetry of the cross-section values relative to the negative and positive values of the anomalous couplings
$\Delta\kappa_{\gamma}$ and $\lambda_{\gamma}$. This is due to the cross terms of the anomalous couplings with SM terms.
It is observed that this asymmetry decreased significantly due to the reduction of the effect of the SM in increasing
center-of-mass energies.

\subsection{Limits on the anomalous couplings $\Delta\kappa_\gamma$ and $\lambda_\gamma$ at the LHeC and the FCC-he}

One of the main purposes of this paper is to determine the best measurements of the anomalous couplings $\Delta\kappa_\gamma$
and $\lambda_\gamma$ at the LHeC and the FCC-he. To carry out this purpose, we adopted a $\chi^ 2$ analysis. The $\chi^ 2$
function for our fit is defined as similar \cite{Mary,Koksal,Gutierrez,Koksal1}:

\begin{equation}
\chi^2(\Delta\kappa_\gamma, \lambda_\gamma )=\Biggl(\frac{\sigma_{SM}-\sigma_{BSM}(\sqrt{s}, \Delta\kappa_\gamma, \lambda_\gamma)}{\sigma_{SM}\sqrt{(\delta_{st})^2+(\delta_{sys})^2}}\Biggr)^2,
\end{equation}

\noindent where $\sigma_{BSM}(\sqrt{s}, \Delta\kappa_\gamma, \lambda_\gamma)$ and $\sigma_{SM}$ are the cross-section in the presence
of beyond SM interactions and in the SM, respectively. $\delta_{st}=\frac{1}{\sqrt{N_{SM}}}$ is the statistical error and $\delta_{sys}$
is the systematic error. The number of events is given by $N_{SM}={\cal L}_{int}\times \sigma_{SM}\times BR(W^{\pm}\to qq', l\nu_l)$, where
${\cal L}_{int}$ is the integrated luminosity and $l=e^-, \mu$. For single $W^-$ production at the LHeC and the FCC-he
we classify their decay products according to the decomposition of $W^-$. In this paper, we consider that the $W^-$ boson decay leptonically
or hadronically for the signal. Thus, we assume that the branching ratios for $W^-$ decays are: $BR(W^- \to q q')=0.674$ for hadronic decays
and $BR(W^- \to l\nu)=0.213$ for light leptonic decays.

We examine in Figs. 5 and 6 the impact of center-of-mass energies $\sqrt{s}=$1.30, 1.98, 7.07, 10 TeV and the
luminosities ${\cal L} = 10, 50, 100, 500, 1000\hspace{0.8mm}{\rm fb^{-1}}$ on the anomalous couplings $\Delta \kappa_\gamma$ and
$\lambda_\gamma$. The expected measurement for both $\Delta\kappa_\gamma$ and $\lambda_\gamma$ is $10^{-2}$ for 7.07, 10\hspace{0.8mm}TeV
and 100, 500, 1000\hspace{0.8mm}${\rm fb^{-1}}$, respectively. For other energy stages of the LHeC, the expected measurements on
$\Delta \kappa_\gamma$ and $\lambda_\gamma$ are an order of magnitude weaker. However, for all the energy and luminosity stages
of the LHeC and the FCC-he the measurements on $\Delta\kappa_\gamma$ and $\lambda_\gamma$ are accessible.

Estimations of the one-parameter limits on the anomalous couplings $\Delta\kappa_\gamma$ and $\lambda_\gamma$ given in Eqs. (8) and (9)
are presented in Table I, where one of the anomalous couplings is fixed to zero. In Table I, we consider the leptonic and hadronic decay
channels of the process $ e^- p \rightarrow e^- \gamma^* p \rightarrow \nu_e W^- p $ at the LHeC with $\sqrt{s}=1.30, 1.98\hspace{0.8mm}{\rm TeV}$
and integrated luminosities ${\cal L}=10, 30, 50, 70, 100 \hspace{0.8mm}{\rm fb^{-1}}$. A similar estimation for the anomalous couplings
$\Delta\kappa_\gamma$ and $\lambda_\gamma$ is presented in Table II, where in this case $\sqrt{s}=7.07, 10\hspace{0.8mm}{\rm TeV}$ at the FCC-he
with integrated luminosities ${\cal L}=100, 300, 500, 700, 1000 \hspace{0.8mm}{\rm fb^{-1}}$, respectively. From these tables, it is clear
that in the leptonic channel the limits on the $\Delta\kappa_\gamma$ and $\lambda_\gamma$ are of the order of magnitude of few times
$10^{-3}$ to $10^{-2}$. However, due to the larger branching ratio, the hadronic channel can improve the constraints by a factor
of two or three with respect to the leptonic channel.

From Table II, our best limits for the anomalous couplings $\Delta\kappa_\gamma$ and $\lambda_\gamma$ at the FCC-he are the following.\\

$i)$ Limits on $\Delta\kappa_\gamma$ and $\lambda_\gamma$ for $\sqrt{s}=7.07\hspace{0.8mm} TeV$, ${\cal L}= 1000 \hspace{0.8mm}{\rm fb^{-1}}$
and $P_{e^-}=0\%$:

\begin{equation}
\Delta\kappa_\gamma =
\begin{array}{ll}
|0.0033|, \hspace{2mm}  & \mbox{$95\%$   C.L.,  \hspace{2mm} leptonic}, \\
|0.0019|, \hspace{2mm}  & \mbox{$95\%$   C.L.,  \hspace{2mm} hadronic},
\end{array}
\end{equation}

\begin{equation}
\lambda_\gamma =
\begin{array}{ll}
|0.0098|, \hspace{2mm}  & \mbox{$95\%$   C.L.,  \hspace{2mm}  leptonic}, \\
|0.0073|, \hspace{2mm}  & \mbox{$95\%$   C.L.,  \hspace{2mm}  hadronic}.
\end{array}
\end{equation}

$ii)$ Limits on $\Delta\kappa_\gamma$ and $\lambda_\gamma$ for $\sqrt{s}=10\hspace{0.8mm} TeV$, ${\cal L}= 1000 \hspace{0.8mm}{\rm fb^{-1}}$
and $P_{e^-}=0\%$:

\begin{equation}
\Delta\kappa_\gamma =
\begin{array}{ll}
|0.0031|, \hspace{2mm}  & \mbox{$95\%$   C.L.,  \hspace{2mm} leptonic}, \\
|0.0017|, \hspace{2mm}  & \mbox{$95\%$   C.L.,  \hspace{2mm} hadronic},
\end{array}
\end{equation}

\begin{equation}
\lambda_\gamma =
\begin{array}{ll}
|0.0071|, \hspace{2mm}  & \mbox{$95\%$   C.L.,  \hspace{2mm}  leptonic}, \\
|0.0053|, \hspace{2mm}  & \mbox{$95\%$   C.L.,  \hspace{2mm}  hadronic}.
\end{array}
\end{equation}

The limits given in Eqs. (21)-(24) are consistent with the corresponding ones of Table I of Ref. \cite{Billur} for the anomalous couplings $\Delta\kappa_\gamma$ and $\lambda_\gamma$.

\subsection{Impact of the polarized electron beam on the cross-section of the process $e^-p \to e^-\gamma^*p \to \nu_eW^- p$ at the LHeC
and the FCC-he}

In the previous sub-sections, the results for the cross-section of the process $e^-p \to e^-\gamma^*p \to \nu_eW^- p$, as well as of
the anomalous parameters $\Delta\kappa_\gamma$ and $\lambda_\gamma$ are presented with unpolarized electron beam. In this sub-section
we discuss the impact of the polarized electron beam in the cross-section and in the anomalous parameters of the aforementioned process.

It is worth noting that a polarized electron beam provides a method to investigate the SM and to diagnose new physics beyond the SM.
Proper selection of the electron beam polarization may, therefore be used to enhance the new physics signal and also to considerably
suppress backgrounds. We select beam polarization as $P_{e^-}=-80\%$  to enhance our physical process. In addition, as we already
mentioned in subsection A, the chiral nature of the weak coupling to fermions results in significant possible enhancements in
$\nu_e W^-$ production, as indicated in Figs. 7 and 8.

Our results for joint variation of the cross-section with the $\Delta\kappa_\gamma$ or $\lambda_\gamma$ couplings are shown in Figs. 7
and 8. In each case, we consider the four center-of-mass energies stages of the FCC-he with their respective integrated luminosities.

The $\sigma(e^-p \to e^-\gamma^*p \to \nu_eW^- p)$ curves as a function of each of the anomalous couplings, setting the other to its SM
value of zero, is shown in Figs. 7 and 8. In this case we consider polarized electron beam with $P_{e^-}=-80\%$. The following
results for the cross-section of the process $\sigma(e^-p \to e^-\gamma^*p \to \nu_eW^- p)$ are obtained: $\sigma(\sqrt{s},
\Delta\kappa_\gamma)= 30\hspace{0.8mm}{\rm pb}$ for $-3\leq\Delta\kappa_\gamma\leq 3$ and $\sigma(\sqrt{s}, \lambda_\gamma)= 10\hspace{0.8mm}{\rm pb}$
for $-2\leq\Delta\kappa_\gamma\leq 2$, in both cases with $\sqrt{s}=10\hspace{0.8mm}{\rm TeV}$. From these figures a difference
of a factor of 4-10 for the minimum and maximum center-of-mass energies of $1.30-10$\hspace{0.8mm}{\rm TeV} is obtained.

\subsection{Impact of the polarized electron beam on the limits of the anomalous couplings $\Delta\kappa_\gamma$ and $\lambda_\gamma$
at the LHeC and the FCC-he}

In this sub-section, we presented a model-independent global fit on the anomalous couplings $\Delta\kappa_\gamma$ and $\lambda_\gamma$.
To carry out this, we made use of the total cross-section for the process $e^-p \to e^-\gamma^*p \to \nu_eW^- p$ in $e^-p$ collisions.
The results of the fit for the four FCC-he energy stages with their respective luminosities are shown in Figs. 9 and 10.

Figs. 9 and 10 show the summary plot illustrating the limits that can be obtained of the process
$e^-p \to e^-\gamma^* p \to \nu_e W^- p$ on the couplings $\Delta \kappa_\gamma$ and $\lambda_\gamma$. We consider the following
center-of-mass energies, luminosities and polarization of the electron beam $\sqrt{s}=1.30, 1.98, 1.07, 10\hspace{0.8mm}{\rm TeV}$,
${\cal L}=10, 50, 100, 500, 1000\hspace{0.8mm}{\rm fb^{-1}}$ and $P_{e^-}=-80\%$, respectively. For comparison, on the same panel
we give the constraints from CMS (grey) Collaborations at the LHC.

To complement our study on the anomalous parameters $\Delta \kappa_\gamma$ and $\lambda_\gamma$ through the process
$e^-p \to e^-\gamma^* p \to \nu_e W^- p$ with polarized electron beam and considering the parameters of the LHeC and
the FCC-he, we give limits for the anomalous couplings of the $W^-$-boson in Tables III and IV. These limits show the
best measurement is compared with the unpolarized case illustrated in Tables I and II.

The following limits are set on the couplings $\Delta \kappa_\gamma$ and $\lambda_\gamma$ at the FCC-he and with polarized
electron beam when one parameter is allowed to vary and the others are set to their SM values of zero.

$i)$ Limits on $\Delta\kappa_\gamma$ and $\lambda_\gamma$ for $\sqrt{s}=7.07\hspace{0.8mm} TeV$, ${\cal L}= 1000 \hspace{0.8mm}{\rm fb^{-1}}$
and $P_{e^-}=-80\%$:

\begin{equation}
\Delta\kappa_\gamma =
\begin{array}{ll}
|0.0025|, \hspace{2mm}  & \mbox{$95\%$   C.L.,  \hspace{2mm} leptonic}, \\
|0.0014|, \hspace{2mm}  & \mbox{$95\%$   C.L.,  \hspace{2mm} hadronic},
\end{array}
\end{equation}

\begin{equation}
\lambda_\gamma =
\begin{array}{ll}
|0.0085|, \hspace{2mm}  & \mbox{$95\%$   C.L.,  \hspace{2mm}  leptonic}, \\
|0.0063|, \hspace{2mm}  & \mbox{$95\%$   C.L.,  \hspace{2mm}  hadronic}.
\end{array}
\end{equation}

$ii)$ Limits on $\Delta\kappa_\gamma$ and $\lambda_\gamma$ for $\sqrt{s}=10\hspace{0.8mm} TeV$, ${\cal L}= 1000 \hspace{0.8mm}{\rm fb^{-1}}$
and $P_{e^-}=-80\%$:

\begin{equation}
\Delta\kappa_\gamma =
\begin{array}{ll}
|0.0023|, \hspace{2mm}  & \mbox{$95\%$   C.L.,  \hspace{2mm} leptonic}, \\
|0.0013|, \hspace{2mm}  & \mbox{$95\%$   C.L.,  \hspace{2mm} hadronic},
\end{array}
\end{equation}

\begin{equation}
\lambda_\gamma =
\begin{array}{ll}
|0.0061|, \hspace{2mm}  & \mbox{$95\%$   C.L.,  \hspace{2mm}  leptonic}, \\
|0.0046|, \hspace{2mm}  & \mbox{$95\%$   C.L.,  \hspace{2mm}  hadronic}.
\end{array}
\end{equation}

A direct comparison of the results shown in the Eqs. (21)-(24) for the unpolarized case and Eqs. (25)-(28) for the case with
polarized electron beam for the anomalous couplings $\Delta\kappa_\gamma$ and $\lambda_\gamma$ clearly shows that the polarized
electron beam effect translates into a factor of 1.35 in the measurement of $\Delta\kappa_\gamma$ and $\lambda_\gamma$.

On the other hand, it is appropriate to mention that the limits shown in Tables I-IV are competitive with the experimental
and phenomenological limits obtained by the ATLAS, CMS, CDF, D0, ALEP, DELPHI, L3 and OPAL Collaborations, as well as by
the ILC and the CEPC which are shown in Table I of Ref. \cite{Billur}.

\section{Conclusions}

The production cross-section of the process $e^-p \to e^-\gamma^* p \to \nu_e W^- p$ in the SM is $1.50\hspace{0.8mm}{\rm pb}$
in the case of unpolarized electron beam and of $3\hspace{0.8mm}{\rm pb}$ for the case of polarized electron beam with
$\sqrt{s}=10\hspace{0.8mm}{\rm TeV}$ as is shown in Figs. 3, 4 and 7, 8. In addition, one can see the
$\sigma(e^-p \to e^-\gamma^* p \to \nu_e W^- p)$ increases monotonically with $\Delta\kappa_\gamma$ and the absolute value
of $\lambda_\gamma$ within the parameter region allowed by current experiments, this is enough to probe anomalous triple
gauge couplings contributions. These couplings could reach ${\cal O}(10^{-3})$ when ${\cal L}=1000\hspace{0.8mm}{\rm fb^{-1}}$.

From the results in Tables I-IV and Figs. 3-10, we could see a significant improvement in the measurement for $\Delta\kappa_\gamma$
and $\lambda_\gamma$ compared to the present ATLAS, CMS, CDF, D0, ALEP, DELPHI, L3 and OPAL collaborations bounds (see Table I
of Ref. \cite{Billur}).

We have presented new searches of anomalous $W^+W^-\gamma$ trilinear gauge boson couplings from
$e^-p \to e^-\gamma^* p \to \nu_e W^- p$ channel analyzing $(10-1000)\hspace{0.8mm}{\rm fb^{-1}}$ of integrated luminosities
and center-of-mass energies $\sqrt{s}=1.30, 1.98, 7.07, 10\hspace{0.8mm}{\rm TeV}$, respectively. We set model-independent limits
on anomalous triple gauge couplings $\Delta\kappa_\gamma$ and $\lambda_\gamma$ for the final states $\nu_e W^- $ at the $95\%$ C.L.:
$\Delta\kappa_\gamma = \pm 0.0017$, $\lambda_\gamma = \pm 0.0053$ (unpolarized electron beam) and $\Delta\kappa_\gamma = \pm 0.0013$,
$\lambda_\gamma = \pm 0.0046$ (polarized electron beam) with $\sqrt{s}=10\hspace{0.8mm}{\rm TeV}$ and ${\cal L}=1000\hspace{0.8mm}{\rm fb^{-1}}$
for both cases. The $W^-$-boson is identified through the hadronic decays channel. In addition, it is worth mentioning that the impact
of the polarized electron beam translates into a factor of 1.35 with respect to the non-polarized case (see Tables I-IV).

In conclusion, the measurement of the $e^-p \to e^-\gamma^* p \to \nu_e W^- p$ channel at the LHeC and the FCC-he would provide a promising
opportunity to probe the anomalous couplings $\Delta\kappa_\gamma$ and $\lambda_\gamma$ without the complications of other couplings especially
QCD backgrounds, and therefore improve our knowledge of the gauge sector. Furthermore, for future measurement of $\Delta\kappa_\gamma$ and
$\lambda_\gamma$, we expect complementarily studies with different electron beam polarizations, as well as a more realistic detector-level
analysis will be very useful.

\begin{table}
\caption{Estimations of the $95\%$\hspace{0.8mm}C.L. prospects for the anomalous couplings $\Delta \kappa_\gamma$ and $\lambda_\gamma$
in the leptonic and hadronic decay channels of the process $ e^- p \rightarrow e^- \gamma^* p \rightarrow \nu_e W^- p $ at the LHeC with
$\sqrt{s}=1.30, 1.98\hspace{0.8mm}{\rm TeV}$ and integrated luminosities of ${\cal L}=10, 30, 50, 70, 100 \hspace{0.8mm}{\rm fb^{-1}}$.
All the limits for $\Delta \kappa_\gamma$($\lambda_\gamma$) are obtained while the other coupling is fixed to their SM value of zero.}
\begin{center}
\begin{tabular}{|c|c|c|c|c|c|}
\hline\hline
 \multicolumn{2}{|c|}{95$\%$ C.L.} & \multicolumn{2}{c}{$\sqrt{s}=$ 1.30 TeV} & \multicolumn{2}{|c|}{$\sqrt{s}=$ 1.98 TeV} \\
\hline
&  &  \multicolumn{4}{c|}{Channel} \\
\cline{3-6}
& ${\cal L} \, (fb^{-1})$  & \hspace{0.8cm} Leptonic \hspace{0.8cm} & \hspace{0.8cm} Hadronic \hspace{0.8cm} & \hspace{0.8cm} Leptonic \hspace{0.8cm} & \hspace{0.8cm} Hadronic \hspace{0.8cm} \\
\hline
\cline{1-6}
 & 10  &  [-0.0607, 0.0571] & [-0.0336, 0.0325] & [-0.0499, 0.0473] & [-0.0277, 0.0269] \\
 & 30  &  [-0.0345, 0.0333] & [-0.0192, 0.0189] & [-0.0285, 0.0276] & [-0.0159, 0.0156] \\
\hspace{0.2cm} $\Delta \kappa_\gamma$ \hspace{0.2cm}
 & 50  &  [-0.0266, 0.0259] & [-0.0149, 0.0146] & [-0.0220, 0.0214] & [-0.0123, 0.0121] \\
 & 70  &  [-0.0224, 0.0219] & [-0.0125, 0.0124] & [-0.0185, 0.0181] & [-0.0103, 0.0102] \\
 & 100 &  [-0.0187, 0.0184] & [-0.0105, 0.0104] & [-0.0155, 0.0152] & [-0.0086, 0.0086] \\
\hline
 & 10  &  [-0.1546, 0.1546] & [-0.1159, 0.1159] & [-0.1022, 0.1022] & [-0.0766, 0.0766] \\
 & 30  &  [-0.1175, 0.1175] & [-0.0881, 0.0881] & [-0.0777, 0.0777] & [-0.0582, 0.0582] \\
$\lambda_\gamma$
 & 50  &  [-0.1034, 0.1034] & [-0.0775, 0.0775] & [-0.0684, 0.0684] & [-0.0512, 0.0512] \\
 & 70  &  [-0.0950, 0.0950] & [-0.0712, 0.0712] & [-0.0628, 0.0628] & [-0.0471, 0.0471] \\
 & 100 &  [-0.0869, 0.0869] & [-0.0652, 0.0652] & [-0.0575, 0.0575] & [-0.0431, 0.0431] \\
\hline\hline
\end{tabular}
\end{center}
\end{table}

\begin{table}
\caption{Estimations of the $95\%$\hspace{0.8mm}C.L. prospects for the anomalous couplings $\Delta \kappa_\gamma$ and $\lambda_\gamma$
in the leptonic and hadronic decay channels of the process $ e^- p \rightarrow e^- \gamma^* p \rightarrow \nu_e W^- p $ at the FCC-he with
$\sqrt{s}=7.07, 10\hspace{0.8mm}{\rm TeV}$ and integrated luminosities of ${\cal L}=100, 300, 500, 700, 1000 \hspace{0.8mm}{\rm fb^{-1}}$.
All the limits for $\Delta \kappa_\gamma$($\lambda_\gamma$) are obtained while the other coupling is fixed to their SM value of zero.}
\begin{center}
\begin{tabular}{|c|c|c|c|c|c|}
\hline\hline
 \multicolumn{2}{|c|}{95$\%$ C.L.} & \multicolumn{2}{c}{$\sqrt{s}=$ 7.07 TeV} & \multicolumn{2}{|c|}{$\sqrt{s}=$ 10 TeV} \\
\hline
&  &  \multicolumn{4}{c|} {Channel}  \\
\cline{3-6}
& ${\cal L} \, (fb^{-1})$ & \hspace{0.8cm} Leptonic \hspace{0.8cm} & \hspace{0.8cm} Hadronic \hspace{0.8cm} & \hspace{0.8cm} Leptonic \hspace{0.8cm} & \hspace{0.8cm} Hadronic \hspace{0.8cm} \\
\hline
\cline{1-6}
 & 100  &  [-0.0107, 0.0106] & [-0.0059, 0.0059] & [-0.0100, 0.0099] & [-0.0055, 0.0055] \\
 & 300  &  [-0.0061, 0.0061] & [-0.0034, 0.0034] & [-0.0057, 0.0057] & [-0.0032, 0.0032] \\
\hspace{0.2cm} $\Delta \kappa_\gamma$ \hspace{0.2cm}
 & 500  &  [-0.0047, 0.0047] & [-0.0026, 0.0026] & [-0.0044, 0.0044] & [-0.0025, 0.0025] \\
 & 700  &  [-0.0040, 0.0040] & [-0.0022, 0.0022] & [-0.0037, 0.0037] & [-0.0021, 0.0021] \\
 & 1000 &  [-0.0033, 0.0033] & [-0.0019, 0.0019] & [-0.0031, 0.0031] & [-0.0017, 0.0017] \\
\hline
 & 100  &  [-0.0175, 0.0175] & [-0.0131, 0.0131] & [-0.0127, 0.0127] & [-0.0095, 0.0095] \\
 & 300  &  [-0.0133, 0.0133] & [-0.0099, 0.0099] & [-0.0096, 0.0096] & [-0.0072, 0.0072] \\
$\lambda_\gamma$
 & 500  &  [-0.0117, 0.0117] & [-0.0088, 0.0088] & [-0.0085, 0.0085] & [-0.0063, 0.0063] \\
 & 700  &  [-0.0107, 0.0107] & [-0.0080, 0.0080] & [-0.0078, 0.0078] & [-0.0058, 0.0058] \\
 & 1000 &  [-0.0098, 0.0098] & [-0.0073, 0.0073] & [-0.0071, 0.0071] & [-0.0053, 0.0053] \\
\hline\hline
\end{tabular}
\end{center}
\end{table}

\begin{table}
\caption{Estimations of the $95\%$\hspace{0.8mm}C.L. prospects for the anomalous couplings $\Delta \kappa_\gamma$ and $\lambda_\gamma$
in the leptonic and hadronic decay channels of the process $ e^- p \rightarrow e^- \gamma^* p \rightarrow \nu_e W^- p $ at the LHeC with
$\sqrt{s}=1.30, 1.98\hspace{0.8mm}{\rm TeV}$ and integrated luminosities of ${\cal L}=10, 30, 50, 70, 100 \hspace{0.8mm}{\rm fb^{-1}}$.
All the limits for $\Delta \kappa_\gamma$($\lambda_\gamma$) are obtained while the other coupling is fixed to their SM value of zero.
We considered polarized electron beam with $P_{e}=-80\%$.}
\begin{center}
\begin{tabular}{|c|c|c|c|c|c|}
\hline\hline
 \multicolumn{2}{|c|}{95$\%$ C.L.} & \multicolumn{2}{c}{$\sqrt{s}=$ 1.30 TeV} & \multicolumn{2}{|c|}{$\sqrt{s}=$ 1.98 TeV} \\
\hline
&  &  \multicolumn{4}{c|} {Channel}  \\
\cline{3-6}
& ${\cal L} \, (fb^{-1})$  & \hspace{0.8cm} Leptonic \hspace{0.8cm} & \hspace{0.8cm} Hadronic \hspace{0.8cm} & \hspace{0.8cm} Leptonic \hspace{0.8cm} & \hspace{0.8cm} Hadronic\hspace{0.8cm} \\
\hline
\cline{1-6}
 & 10  &  [-0.0449, 0.0429] & [-0.0249, 0.0243] & [-0.0369, 0.0355] & [-0.0205, 0.0201] \\
 & 30  &  [-0.0256, 0.0250] & [-0.0143, 0.0141] & [-0.0211, 0.0206] & [-0.0118, 0.0117] \\
\hspace{0.2cm} $\Delta \kappa_\gamma$ \hspace{0.2cm}
 & 50  &  [-0.0198, 0.0194] & [-0.0110, 0.0109] & [-0.0163, 0.0160] & [-0.0091, 0.0090] \\
 & 70  &  [-0.0167, 0.0164] & [-0.0093, 0.0092] & [-0.0137, 0.0135] & [-0.0076, 0.0076] \\
 & 100 &  [-0.0139, 0.0138] & [-0.0078, 0.0077] & [-0.0113, 0.0113] & [-0.0064, 0.0064] \\
\hline
 & 10  &  [-0.1335, 0.1335] & [-0.1001, 0.1001] & [-0.0882, 0.0882] & [-0.0661, 0.0661] \\
 & 30  &  [-0.1014, 0.1014] & [-0.0760, 0.0760] & [-0.0670, 0.0670] & [-0.0502, 0.0502] \\
$\lambda_\gamma$
 & 50  &  [-0.0892, 0.0892] & [-0.0669, 0.0669] & [-0.0590, 0.0590] & [-0.0442, 0.0442] \\
 & 70  &  [-0.0820, 0.0820] & [-0.0615, 0.0615] & [-0.0542, 0.0542] & [-0.0406, 0.0406] \\
 & 100 &  [-0.0750, 0.0750] & [-0.0562, 0.0562] & [-0.0496, 0.0496] & [-0.0372, 0.0372] \\
\hline\hline
\end{tabular}
\end{center}
\end{table}

\begin{table}
\caption{Estimations of the $95\%$\hspace{0.8mm}C.L. prospects for the anomalous couplings $\Delta \kappa_\gamma$ and $\lambda_\gamma$
in the leptonic and hadronic decay channels of the process $ e^- p \rightarrow e^- \gamma^* p \rightarrow \nu_e W^- p $ at the FCC-he with
$\sqrt{s}=7.07, 10\hspace{0.8mm}{\rm TeV}$ and integrated luminosities of ${\cal L}=100, 300, 500, 700, 1000 \hspace{0.8mm}{\rm fb^{-1}}$.
All the limits for $\Delta \kappa_\gamma$($\lambda_\gamma$) are obtained while the other coupling is fixed to their SM value of zero.
We considered polarized electron beam with $P_{e}=-80\%$.}
\begin{center}
\begin{tabular}{|c|c|c|c|c|c|}
\hline\hline
 \multicolumn{2}{|c|}{95$\%$ C.L.} & \multicolumn{2}{c}{$\sqrt{s}=$ 7.07 TeV} & \multicolumn{2}{|c|}{$\sqrt{s}=$ 10 TeV} \\
\hline
&  &  \multicolumn{4}{c|} {Channel}  \\
\cline{3-6}
& ${\cal L} \, (fb^{-1})$  & \hspace{0.8cm} Leptonic \hspace{0.8cm} & \hspace{0.8cm} Hadronic \hspace{0.8cm} & \hspace{0.8cm} Leptonic \hspace{0.8cm} & \hspace{0.8cm} Hadronic \hspace{0.8cm} \\
\hline
\cline{1-6}
 & 100  &  [-0.0080, 0.0079] & [-0.0044, 0.0044] & [-0.0074, 0.0074] & [-0.0041, 0.0041] \\
 & 300  &  [-0.0045, 0.0045] & [-0.0025, 0.0025] & [-0.0042, 0.0042] & [-0.0024, 0.0024] \\
\hspace{0.2cm} $\Delta \kappa_\gamma$ \hspace{0.2cm}
 & 500  &  [-0.0035, 0.0035] & [-0.0020, 0.0020] & [-0.0033, 0.0033] & [-0.0018, 0.0018] \\
 & 700  &  [-0.0030, 0.0030] & [-0.0016, 0.0016] & [-0.0028, 0.0028] & [-0.0015, 0.0015] \\
 & 1000 &  [-0.0025, 0.0025] & [-0.0014, 0.0014] & [-0.0023, 0.0023] & [-0.0013, 0.0013] \\
\hline
 & 100  &  [-0.0151, 0.0151] & [-0.0113, 0.0113] & [-0.0110, 0.0110] & [-0.0082, 0.0082] \\
 & 300  &  [-0.0115, 0.0115] & [-0.0086, 0.0086] & [-0.0083, 0.0083] & [-0.0062, 0.0062] \\
$\lambda_\gamma$
 & 500  &  [-0.0101, 0.0101] & [-0.0075, 0.0075] & [-0.0073, 0.0073] & [-0.0055, 0.0055] \\
 & 700  &  [-0.0093, 0.0093] & [-0.0069, 0.0069] & [-0.0067, 0.0067] & [-0.0050, 0.0050] \\
 & 1000 &  [-0.0085, 0.0085] & [-0.0063, 0.0063] & [-0.0061, 0.0061] & [-0.0046, 0.0046] \\
\hline\hline
\end{tabular}
\end{center}
\end{table}


\newpage

\vspace*{5mm}

\begin{center}
{\bf Acknowledgments}
\end{center}

A. G. R. and M. A. H. R. thank SNI and PROFOCIE (M\'exico).

\vspace{2cm}


\pagebreak

\begin{figure}[t]
\centerline{\scalebox{0.8}{\includegraphics{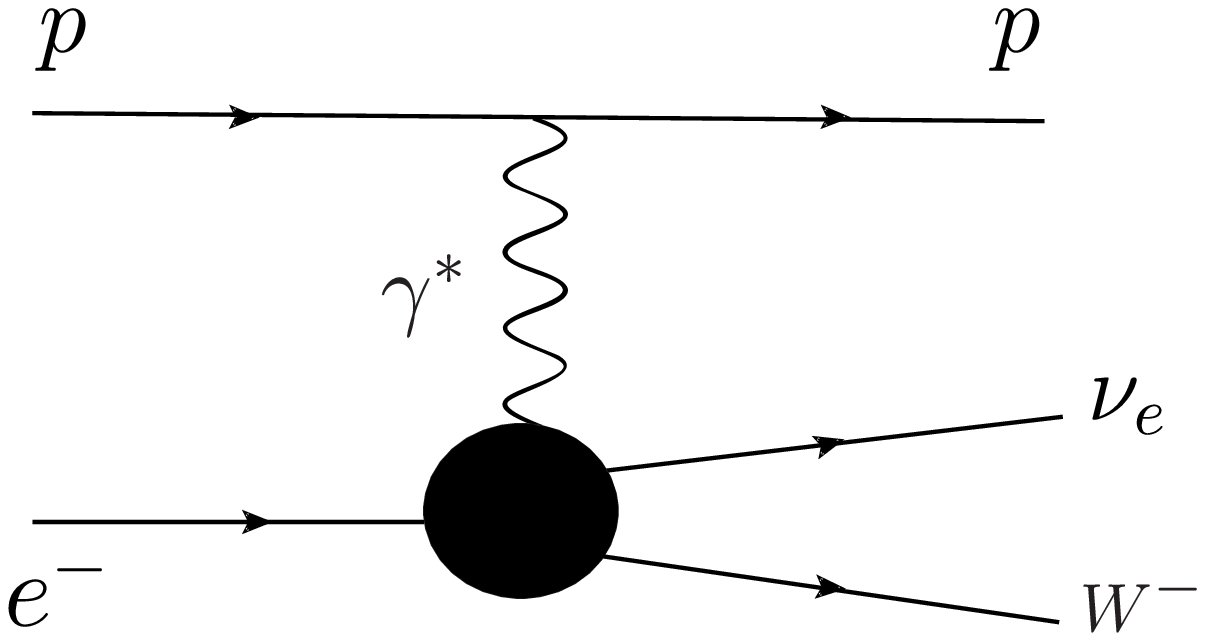}}}
\caption{ \label{fig:gamma1} A schematic diagram for the process
$e^-p \to e^-\gamma^*p \to \nu_eW^- p$.}
\label{Fig.1}
\end{figure}

\begin{figure}[t]
\centerline{\scalebox{0.8}{\includegraphics{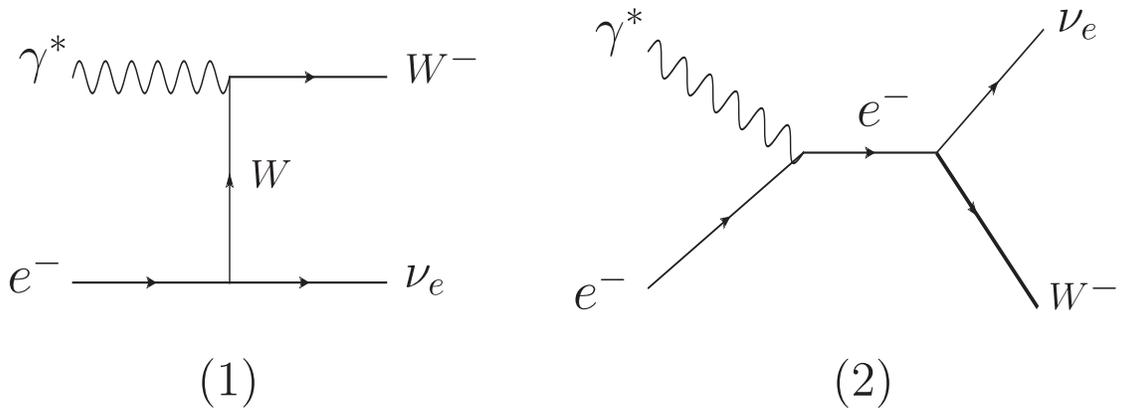}}}
\caption{ \label{fig:gamma2} Feynman diagrams contributing to the subprocess
$\gamma^*e^- \to \nu_e W^-$.}
\label{Fig.2}
\end{figure}

\begin{figure}[t]
\centerline{\scalebox{1.4}{\includegraphics{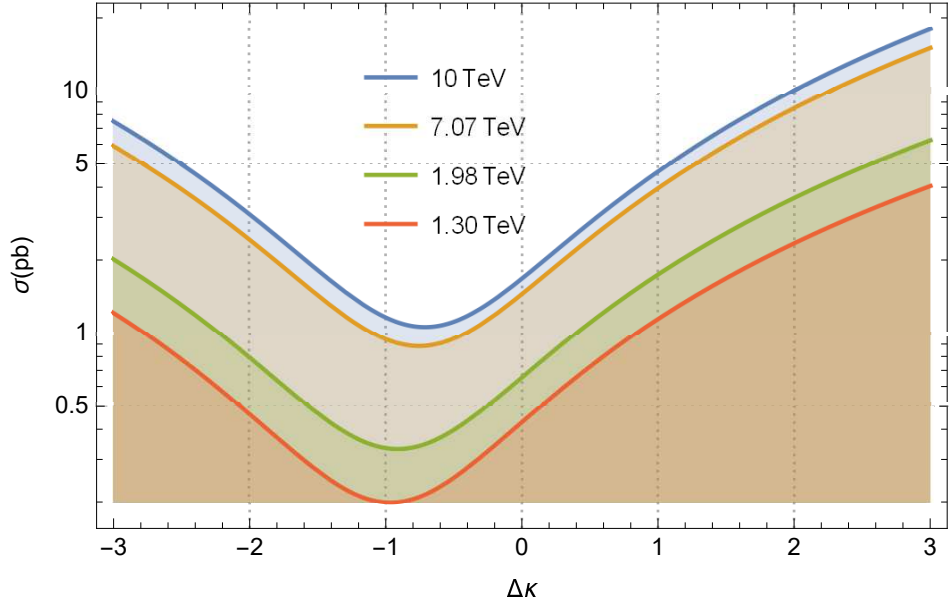}}}
\caption{ The total cross sections of the process
$e^-p \to e^-\gamma^* p \to \nu_e W^- p$ as a function of $\Delta \kappa_\gamma$
for center-of-mass energies of $\sqrt{s}=1.30, 1.98, 7.07, 10\hspace{0.8mm}{\rm TeV}$
at the LHeC and the FCC-he.}
\label{Fig.3}
\end{figure}

\begin{figure}[t]
\centerline{\scalebox{1.4}{\includegraphics{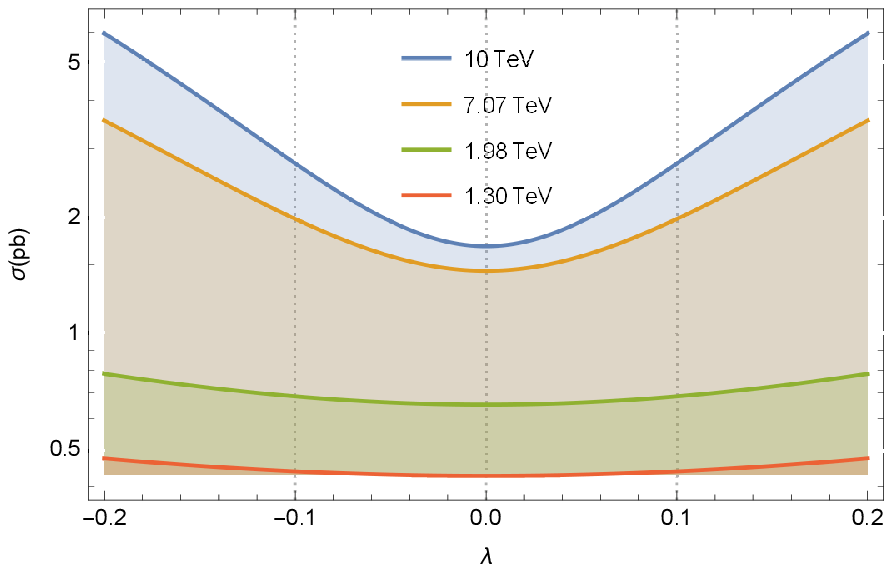}}}
\caption{ Same as in Fig. 3, but for $\lambda_\gamma$.}
\label{Fig.4}
\end{figure}

\begin{figure}[t]
\centerline{\scalebox{0.95}{\includegraphics{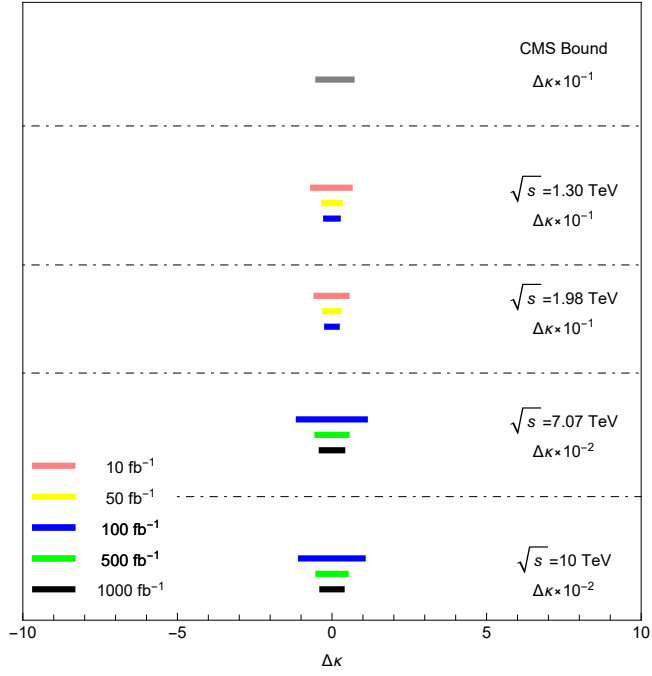}}}
\caption{ \label{fig:gamma1} Comparison of precisions at the LHeC and the FCC-he to the anomalous coupling
$\Delta\kappa_\gamma$ for center-of-mass energies $\sqrt{s}=1.30, 1.98, 1.07, 10\hspace{0.8mm}{\rm TeV}$
and luminosities ${\cal L}=10, 50, 100, 500, 1000\hspace{0.8mm}{\rm fb^{-1}}$. We consider the process
$e^-p \to e^-\gamma^* p \to \nu_e W^- p$. We include the CMS bound.}
\label{Fig.5}
\end{figure}

\begin{figure}[t]
\centerline{\scalebox{0.95}{\includegraphics{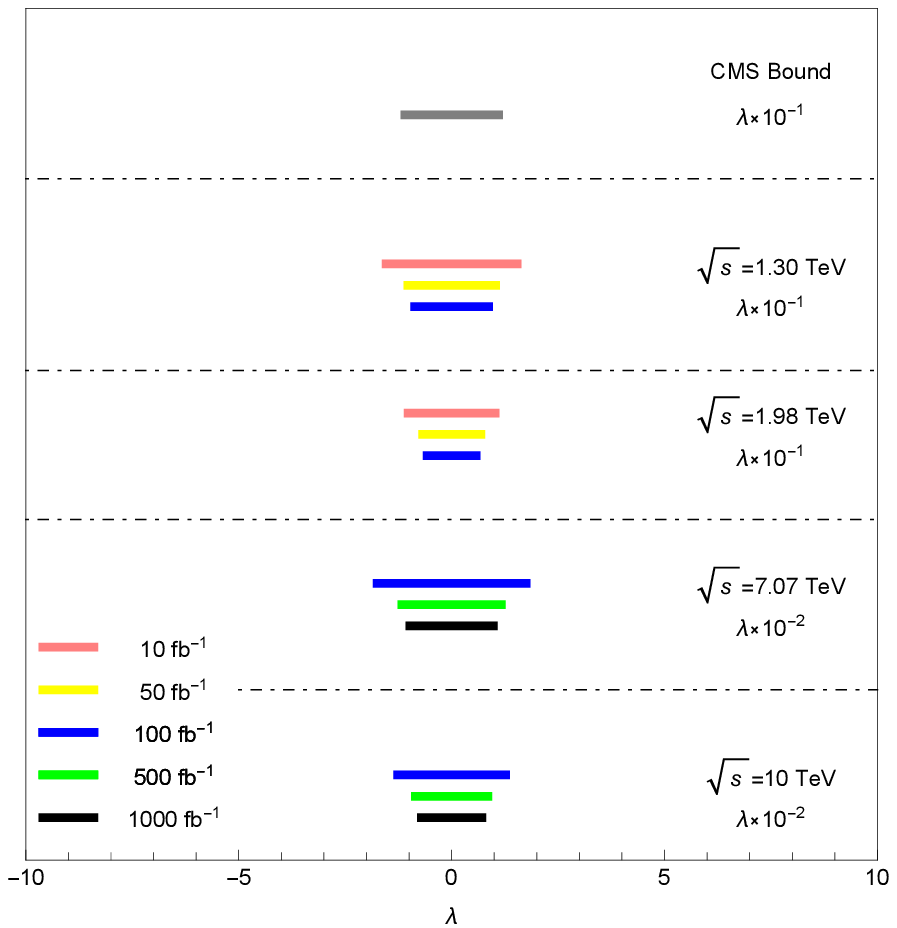}}}
\caption{ \label{fig:gamma2} Same as in Fig. 5, but for $\lambda_\gamma$.}
\label{Fig.6}
\end{figure}

\begin{figure}[t]
\centerline{\scalebox{1.4}{\includegraphics{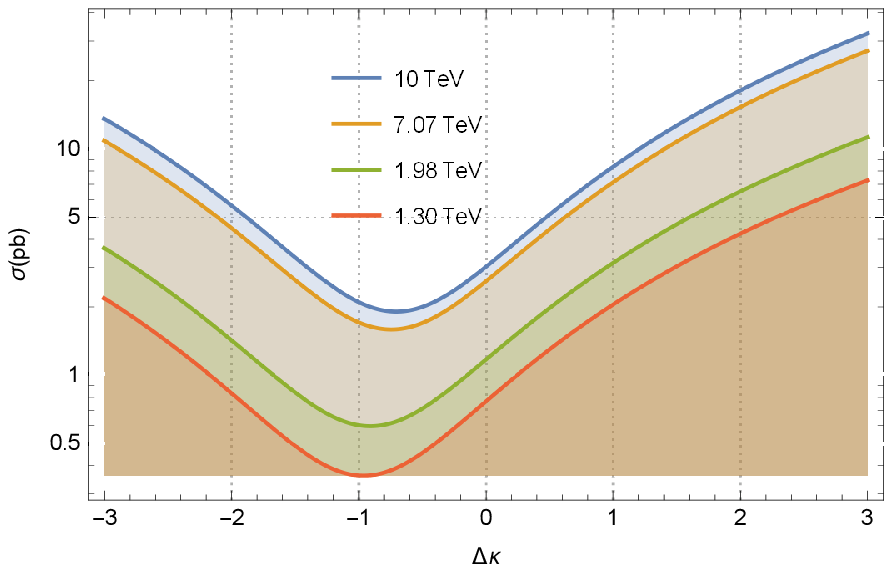}}}
\caption{ \label{fig:gamma1x} Same as in Fig. 3, but with polarized electron beam.}
\label{Fig.7}
\end{figure}

\begin{figure}[t]
\centerline{\scalebox{1.4}{\includegraphics{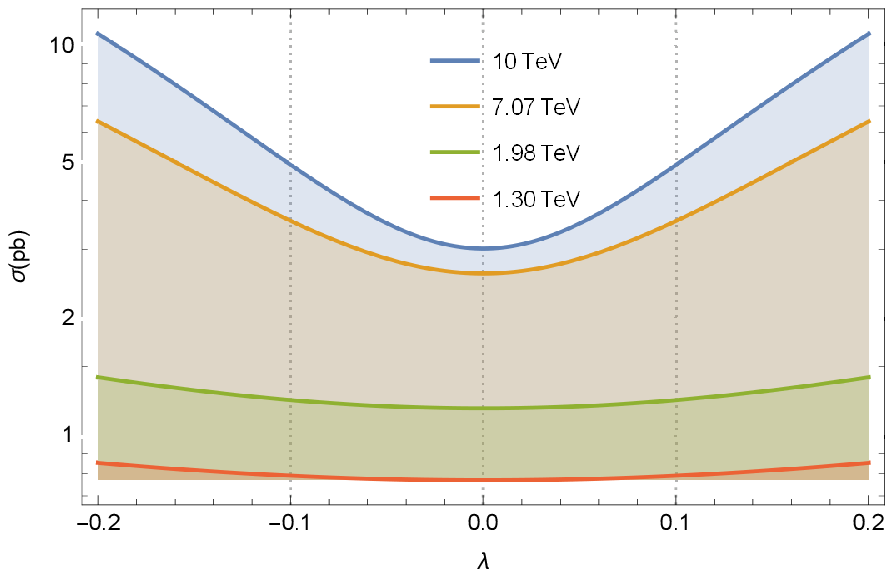}}}
\caption{ \label{fig:gamma2x} Same as in Fig. 4, but with polarized electron beam.}
\label{Fig.8}
\end{figure}

\begin{figure}[t]
\centerline{\scalebox{1}{\includegraphics{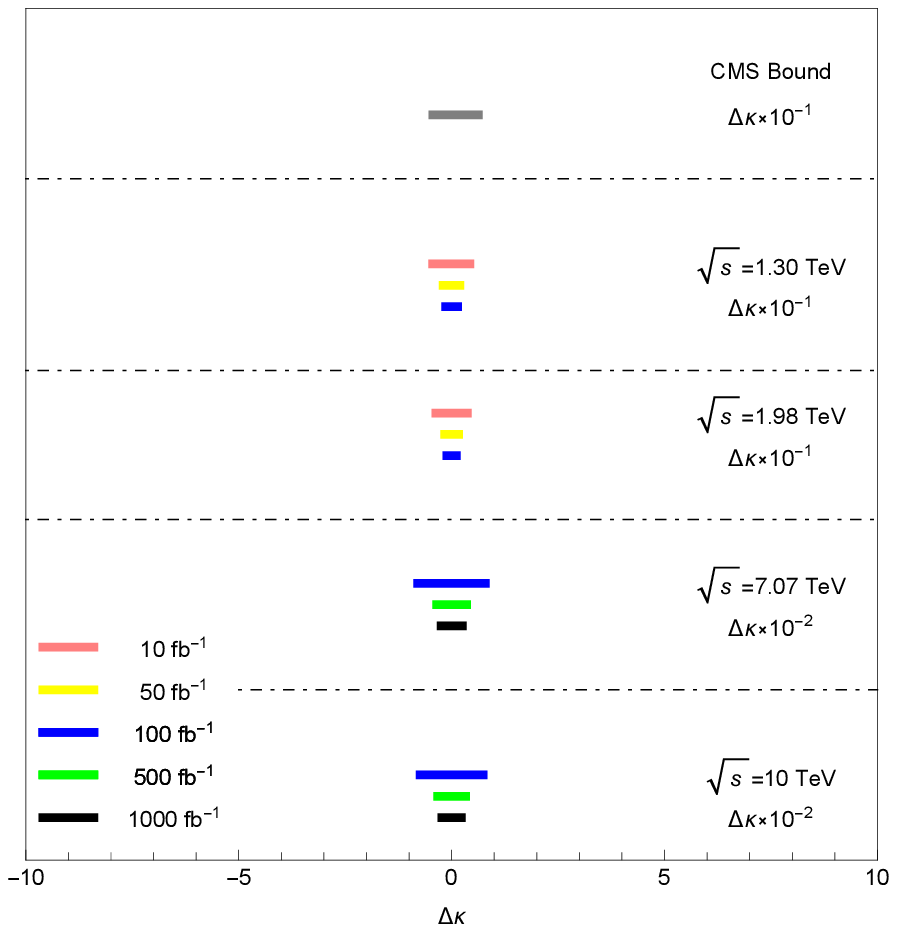}}}
\caption{ \label{fig:gamma15} Same as in Fig. 5, but with polarized electron beam.}
\label{Fig.6}
\end{figure}

\begin{figure}[t]
\centerline{\scalebox{1}{\includegraphics{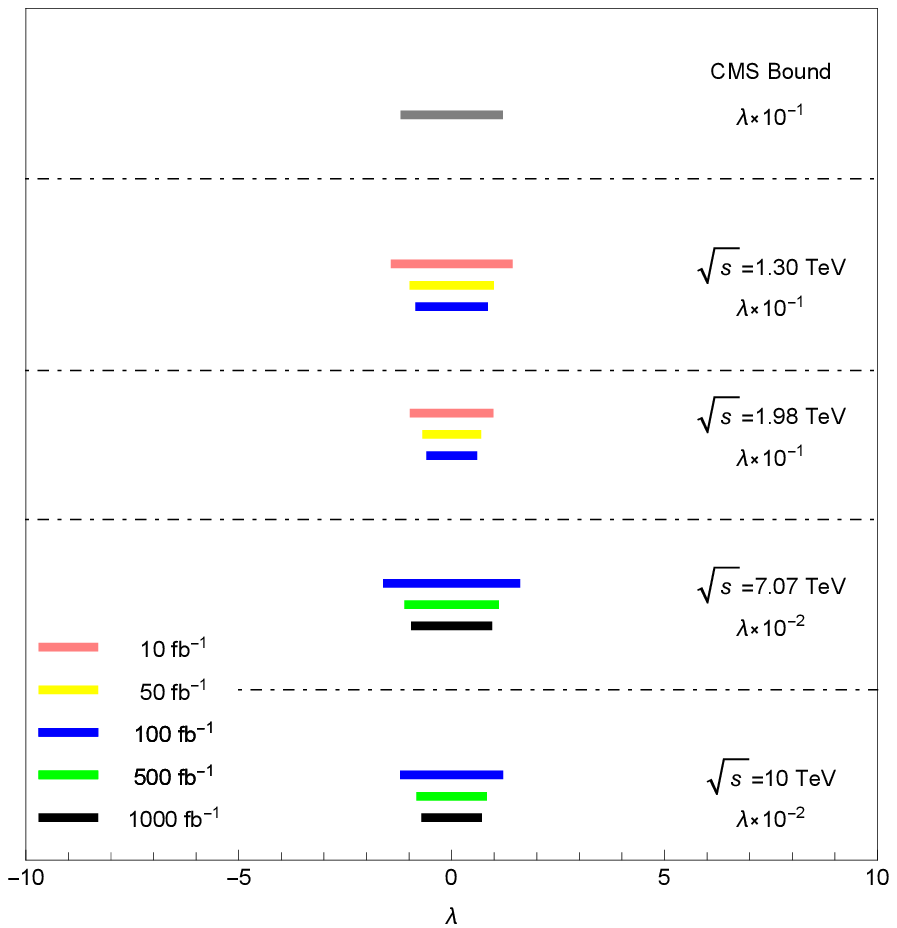}}}
\caption{ \label{fig:gamma6} Same as in Fig. 6, but with polarized electron beam.}
\label{Fig.7}
\end{figure}

\end{document}